# Characterization of Hydroxyls in Surface Oxide of Superconducting Tantalum and Their Mitigation in Quantum Circuits

Ekta Bhatia[1], Nicholas Pieniazek[1], Aleksandra Biedron[1], Sandra Schujman[1], Hunter Frost[2], Zhihao Xiao[1], Jakub Nalaskowski[1], Kevin Musick[1], Thomas Murray[1], Satyavolu Papa Rao[1]

[1]*NY Creates, Albany, NY 12203, USA*

[2]*College of Nanotechnology, Sci. & Eng., University at Albany (SUNY), Albany, NY 12203, USA*

Recently, tantalum (Ta) has gained attention in superconducting quantum circuits due to the longer coherence times achieved when replacing niobium (Nb) in capacitor pads. Previous literature shows that surface oxides that form upon ambient exposure on superconducting metals such as Ta, Al, and Nb host two-level system (TLS) defects, which are a leading source of microwave loss and decoherence. While the surface oxides of Nb and Al have been extensively studied, Ta oxides remain less well understood. Using secondary ion mass spectrometry of α-Ta films deposited at 300 mm wafer scale, we show for the first time that hydroxyls accumulate in the Ta suboxide region above the underlying Ta. Angle-resolved X-ray photoelectron spectroscopy shows that the surface region is dominated by $Ta_2O_5$, with sub-stoichiometric $TaO_x$ present in between the $Ta_2O_5$ and underlying Ta. The thickness of the tantalum oxide is confirmed by transmission electron microscopy. We demonstrate that [OH] incorporation can be suppressed by replacing the native oxide with an oxide formed during chemical mechanical planarization of α-Ta films. Our findings support the hypothesis that TLS defects are non-uniform within the oxide thickness and suggest hydroxyls as a probable molecular origin of these loss channels. Furthermore, we show the feasibility of plasma nitridization as a method to decrease hydroxyl loading on α-Ta surfaces. The modulation of hydroxyl content through surface engineering of α-Ta can enable the fabrication of more robust, high-coherence superconducting quantum circuits by addressing a potential TLS source.

## I. INTRODUCTION

Superconducting qubits have emerged as a leading architecture for quantum information processing, enabling advances in quantum error correction, studies of many-body physics, and quantum simulation [1-3]. While steady improvements in fabrication and design have improved device performance, qubit coherence times remain a primary constraint on the scalability of these systems [4,5]. Among the various decoherence mechanisms, dielectric loss arising from two-level system (TLS) defects in surface oxides has been identified as a dominant source of energy relaxation in transmon qubits [5,6]. Recent works have shown that qubits fabricated from Ta can achieve record coherence times [7]. This improvement is attributed in part to Ta's favorable material properties, including thermal robustness, chemical resistance, and a thinner native oxide with less complex chemistry compared to Nb and Al [5,7]. Nevertheless, even in state-of-the-art tantalum devices, losses are largely due to TLS defects residing within the interfaces of tantalum with air [8]. Therefore, it is important to systematically study and control the formation of Ta surface oxides to understand the origin and spatial distribution of TLS defects and to develop strategies that improve coherence and reliability in superconducting quantum devices.

Several approaches have been proposed to mitigate TLS-related losses by modifying surface chemistry, including encapsulation [9, 10, 14], in-situ surface treatments [11], and controlled oxidation [12, 13]. Another strategy for mitigating TLS-induced losses is vacuum annealing to remove the surface oxide. This method has proven effective in superconducting radio-frequency (SRF) cavities [15], but the high temperatures and ultra-high vacuum conditions required are generally incompatible with multilayer or heterojunction quantum devices. An alternative approach is to modify the metal surface chemistry through nitrogen enrichment [16], where the increased interstitial nitrogen hinders oxygen in-diffusion and suppresses the growth of $NbO_x$. In this work, we have used chemical mechanical planarization (CMP), which is an industry-standard process that uses a polishing pad and chemical slurry to remove material from the wafer surface under applied pressure, resulting in a planar and smooth surface. The slurry contains an oxidizing agent that promotes the growth of a chemically-formed surface oxide [13, 17]. Plasma nitridization of the Ta surface to form $TaN_x$, was also studied in this work, to characterize its efficacy as a diffusion barrier to potentially reduce TLS defect formation in industry grade tools on 300 mm wafers.

Using a combination of angle-resolved X-ray photoelectron spectroscopy (ARXPS), Secondary Ion Mass Spectrometry (SIMS), and cross-sectional transmission electron microscopy (TEM), we determine the spatial distribution of tantalum oxidation states and hydroxyl content across the surface and suboxide regions. Our results show that hydroxyls accumulate in the sub-stoichiometric $TaO_x$ layer and that their incorporation can be reduced through CMP and nitridization. These findings provide insights into the spatial location of TLS defects in tantalum-based devices and highlight viable pathways to suppress decoherence mechanisms in superconducting quantum devices.

## II. EXPERIMENTAL DETAILS

### A. Sample preparation

A 3 nm TaN layer was deposited by atomic layer deposition (ALD), immediately followed by physical vapor deposition (PVD) of Ta, in a 300 mm cluster tool capable of *in vacuo* transfer between the two process chambers. The use of ALD TaN as a seed layer enabled the formation of an α-phase Ta film. Four surfaces were studied: (i) native Ta oxide formed by leaving the Ta thin film in ambient air for three weeks [18], (ii) chemically formed oxide via CMP, as described earlier, (iii) plasma nitridization of the Ta surface, followed by immediate capping with Nb metal and (iv) plasma nitridization of the Ta surface followed by exposure to ambient air for three weeks, followed by capping with Nb metal.

**B. Angle-resolved X-ray photoelectron spectroscopy**

ARXPS measurements on coupons (each 10 × 5 $mm^2$) were performed using a PHI Quantera Hybrid XPS system that uses a monochromatic Al Kα X-ray source with photon energy of 1486.6 eV) and a focal spot of size ~200 µm. The XPS data were collected at take-off angles of 30°, 45°, and 70° for Ta 4f, O 1s, N 1s, and C 1s spectra. Spectra were acquired at a pass energy of 55 eV with a step size of 0.05 eV, using an angular acceptance of 5°.

Data analysis was performed with the CasaXPS software package. A Shirley-type background was subtracted from Ta4f spectra over the binding energy window 19 eV to 31 eV. Peak fitting [19] was performed using previously reported binding-energy shifts of tantalum oxides and values from the NIST XPS database [20].

Each Ta 4f spectrum was deconvoluted into spin–orbit doublets, with the separation fixed at 1.91 eV and the area ratio between Ta $4f_{7/2}$ and Ta $4f_{5/2}$ constrained to 4:3. The metallic Ta signal was modeled with a Lorentzian line shape with an asymmetry correction. For Ta oxides ($TaO_{x<2.5}$, $Ta_2O_5$), symmetric Gaussian–Lorentzian (GL(50)) line shapes were used. The full width at half maximum (FWHM) was restricted to 0.5–1.5 eV for the sharp metallic peak and 0.5–2 eV for oxide peaks, following recommended XPS fitting best practices [18].

The Ta 4f spectra were best described by three doublets corresponding to metallic Ta and two oxidation states ($Ta^{x<2.5+}$, $Ta^{5+}$), with $Ta_2O_5$ being the dominant oxide. The analysis confirmed that the chemical shift of the $Ta^{x+}$ oxide peaks relative to the $Ta^0$ metal peak increases monotonically with oxidation state, such that $Ta^{5+}$ has the largest shift, followed by the lower oxidation states [19, 20].

**C. Secondary ion mass spectrometry**

Secondary Ion Mass Spectrometry (SIMS) was performed using a TOF.SIMS 5 instrument (IONTOF GmbH, Münster, Germany). Depth profiling was carried out using dual-beam operation. A 500 eV $Cs^+$ sputter beam progressively removed material in a 600 × 600 $\mu m^2$ area, and secondary ion spectra were acquired with a 25 keV $Bi^{3+}$ analysis beam (rastered over a 200 × 200 $\mu m^2$ region, centered within the sputter crater). The signals were normalized to the total secondary ion counts at each depth to allow for relative comparison of samples prepared with different surface treatments. The peak intensities and areas of different ionic species, such as $OH^-$, $TaOH^-$, $TaN^-$, $Ta_2O_5^-$ and metallic $Ta^-$, were analyzed using IONTOF SurfaceLab 7.4 software.

**D. High-resolution transmission electron microscopy**

To compare the thicknesses of the native, CMP-formed, and nitridization-formed surface oxides and oxynitride, HR-TEM was performed using a ThermoFisher Titan³ G2 80-300 microscope. The instrument was equipped with an XFEG source, a monochromator, and a probe Cs corrector. Imaging was carried out in two-condenser lens TEM mode at 300 kV with a beam current of ~1 nA. A 150 µm C2 aperture and a 70 µm objective aperture were used, and images were taken with a bottom-mounted Gatan Ultrascan detector.

Lamellae for TEM were prepared via focused ion beam (FIB) after a protective carbon cap was deposited on samples to preserve surface morphology and prevent contamination during FIB preparation. Oxide and nitridized layer thicknesses were measured, and interface roughness was qualitatively assessed from high-resolution images.

## III. RESULTS AND DISCUSSION

ARXPS was performed at take-off angles of 30°, 45°, and 70° to investigate the depth-dependent chemical states in tantalum surface oxides. As the take-off angle decreases, the spectra become more surface-sensitive. In Figure 1a, Ta 4f core-level spectra acquired at each angle are shown after normalization of the $Ta^{5+}$ peak intensity at each angle – this enables a direct comparison of angle-dependent changes. The Ta $4f_{7/2}$ and $4f_{5/2}$ peaks corresponding to the fully oxidized state ($Ta^{5+}$ in $Ta_2O_5$) dominate across all angles.

Figure 1b quantifies the peak intensities extracted from panel (a), grouped by chemical state and plotted for each angle. While the $Ta_2O_5$ intensity is held constant across all angles, the relative intensities of $TaO_{x<2.5}$ and metallic Ta increase at higher take-off angles (e.g., 70°) which is consistent with detecting higher Ta signal at deeper depths in bulk Ta. This suggests a stratified oxide structure consisting of a surface layer rich in $Ta_2O_5$, an intermediate suboxide layer, and underlying metallic Ta, as illustrated in the schematic inset of Figure 1b.  This finding is similar to that of prior reports [24]

SIMS was performed to investigate the depth distribution of tantalum oxides in samples with native and chemically formed surface oxides. Figure 2 shows the normalized intensities of $Ta_2O_5$, TaOH, and metallic Ta, plotted as a function of sputter depth for both native and CMP-formed oxides. The SIMS data has been normalized by the total secondary ion yield. It can be seen that the normalized $Ta^{5+}$ signal ($Ta_2O_5$/Total) dominates the near-surface region in both cases, indicating a fully oxidized topmost layer. As sputtering progresses, a decrease in $Ta^{5+}$ and an increase in the TaOH signal is observed, indicating that hydroxyls preferentially accumulate beneath the fully oxidized $Ta_2O_5$ region. The XPS analysis independently confirms that the region at $Ta_2O_5$/Ta interface is dominated by sub-stoichiometric $TaO_x$ with $x < 2.5$. The coincidence of the TaOH/Total peak within this window suggests that hydroxyl related species preferentially accumulate in the sub-oxide region. The hydroxyl signal (TaOH/Total) is significantly higher in the native oxide sample compared to the CMP-formed oxide, which suggests greater hydroxyl incorporation and accumulation within the native Ta suboxide region. As we go deeper, the Ta/Total signal increases in the bulk region that shows the transition to the metallic Ta layer for both samples. We postulate that hydrogen or hydroxyl ions diffusing in from the surface find it far easier to bind to the sub-stoichiometric Ta oxide, than to the fully-coordinated Ta atoms present in the Ta2O5 layer at the surface, giving rise to this accumulation in the sub-surface region. A small residual $TaOH^-$/Total signal is also observed in the metallic Ta region; this may reflect hydroxyl species trapped at Ta grain boundaries or interfaces, although an apparent enhancement due to total-yield normalization cannot be excluded because the total ion yield is lower in the metallic Ta region than in the native oxide region. Additionally, it should be noted that SIMS analysis of samples taken from the wafer edge show results that mirror those shown in Figure 2 (which is for a sample taken from the wafer center).

SIMS results shown in Fig. 2 support the hypothesis found in the literature that TLS defects are not uniformly distributed throughout the oxide [8]. Previous studies of aluminum oxides [21] have implicated the rotational freedom of [$OH^-$] bonds as contributing to TLS. It is plausible that hydroxyl groups may serve as a molecular origin of TLS in tantalum surface oxides and these are most prominent in $TaO_x$ region at Ta/$Ta_2O_5$ interface. This behavior also mirrors findings in niobium films, where hydrogen-containing defects were found in the suboxide region near the Nb/$Nb_2O_5$ interface [13]. The observed decrease in the hydroxyl-related SIMS signal content in CMP-formed tantalum oxide suggests CMP as a viable surface-engineering strategy to help reduce TLS-related decoherence in superconducting qubits.

The thickness of surface layers formed by native oxidation, CMP, and plasma nitridization of Ta was quantified by cross-sectional TEM (Fig. 3). Figure 3a shows a ~3 nm thick native oxide formed on Ta following ambient exposure for 3 weeks. In contrast, the CMP-treated sample (Fig. 3b) has a surface oxide layer of comparable thickness – but on that is present on a planarized Ta surface, without the topography immediately obvisou in Figure 3(a). This observation is consistent with CMP removing the pre-existing native oxide in the process of planarizing the Ta surface and simultaneously promoting re-oxidation of the freshly exposed Ta surface through slurry chemistry. The planar Ta surface observed in the CMP sample provides an additional advantage for device integration.

Figure 3c shows a nitridized Ta sample, where a distinct nitridized layer forms at the Ta surface due to its reaction with nitrogen plasma. It should be noted that the Ta surface was not subjected to CMP prior to nitridization, explaining the topography that can be observed on the Ta surface. Although TEM contrast alone cannot reliably distinguish compositional or density changes, the presence of a continuous near-surface layer suggests that nitrogen incorporation modifies the near-surface composition and may act as a barrier to oxygen and hydroxyl diffusion. The conformal morphology indicates that nitridization proceeds without altering the underlying Ta roughness.

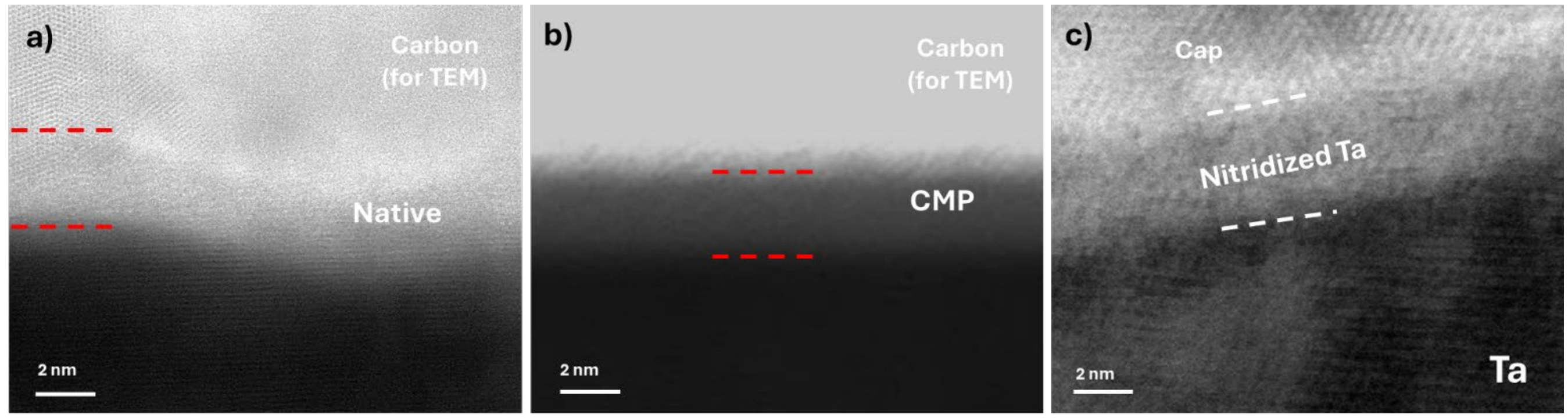

**FIG. 3.** Cross-sectional HR-TEM images of surface oxides on tantalum under different processing conditions: (a) Native oxide formed via ambient exposure. (b) CMP-processed sample shows a more planar surface and similar thickness as native oxide. (c) Nitridized Ta surface shows a nitridized layer under a protective cap deposited after 3 weeks of ambient exposure.

To evaluate the impact of surface nitridization followed by air exposure on hydroxyl incorporation in Ta suboxide region, SIMS depth profiling was performed on a nitridized sample with the same air exposure duration prior to Nb capping. Figure 4 shows the depth distribution of $TaOH^-$ and $TaN^-$ secondary ion signals normalized to the total counts for a nitridized sample capped after 3 weeks of ambient exposure and its comparison with native oxide sample exposed to air for 3 weeks. For the nitridized sample, the $TaOH^-$ signal is suppressed compared to samples exposed to air for three weeks prior to capping (red and blue traces). The formation of Ta-N bonds on the near surface region by plasma nitridization is thought to slow the diffusion kinetics for hydroxyls, and also reduce available sites for bonding. The $TaOH^-$ peak resides near the interface between the capping layer and the underlying nitride. This indicates that hydroxyls are introduced post-nitridization through environmental interaction, whereas the $TaN^-$ signal marks the extent of the nitrided region, with a transition to the underlying Ta film.

Figure 5 quantifies the total maximum $TaOH^-$/Total signal within the hydroxyl-rich region (in the $TaO_x$ region) across different surface treatments and air exposure conditions. The native oxide with 3 weeks of air exposure exhibits the highest hydroxyl content, followed by the CMPed oxide exposed to air for the same time interval. The nitridized sample with the air exposure for the same time interval shows the lowest $TaOH^-$ formation. These findings support the hypothesis that $OH^-$ form during ambient exposure and become localized at or near the metal–oxynitride interface and surface nitridization can modulate hydroxyl incorporation in a way that is expected to influence TLS-related decoherence in superconducting quantum circuits.

## IV. CONCLUSIONS

In this work, we combined ARXPS, SIMS, and TEM to study the depth-dependent chemical structure of native Ta oxides. These measurements show that the surface of the native oxide is dominated by $Ta_2O_5$, with a substoichiometric $TaO_{x<2.5}$ region located near the Ta/oxide interface. The depth profiling by SIMS further shows that hydroxyls preferentially accumulate within the suboxide region. This observation is consistent with prior hypotheses that TLS are not uniformly distributed across amorphous surface oxides – with this work providing evidence that hydroxyls accumulate in the sub-stoichiometric oxide region that is located below the $Ta_2O_5$ and just above the metallic Ta. Accordingly, the presence of hydroxyl species in $TaO_{x<2.5}$ is consistent with proposed microscopic models in which OH- and hydrogen-related defects contribute to TLS and dielectric loss, as reported in Nb- and Al-based systems.

To mitigate hydroxyl incorporation in this suboxide region, we used two surface-engineering approaches: CMP and plasma nitridization. CMP produces an oxide with reduced maximum TaOH- content, a more uniform thickness, and a planar metal–oxide interface compared to the native oxide. In contrast, plasma nitridization creates a nitridized layer on Ta surface which forms an oxynitride layer upon ambient exposure and the hydroxyl incorporation is suppressed compared to native oxide. Future work will quantify the influence of reduced hydroxyl incorporation on TLS loss in resonators and coherence metrics using device integration at 300 mm scale [22, 23].

## ACKNOWLEDGMENTS

This study was supported by the U.S. Department of Energy, Office of Science, National Quantum Information Science Research Centers, Co-design Center for Quantum Advantage (C2QA), under Contract No. DE-SC0012704, including Subcontract No. 390040.

## AUTHOR DECLARATIONS

### Conflict of Interest

E.B. is serving as a Guest Editor for the AVS Quantum Science Special Issue "Selected Papers from the AVS 71 Quantum Mini-Symposium 2025." E.B. had no role in the editorial handling, peer review, or decision-making process for this manuscript. The authors have no other conflicts to disclose.

## DATA AVAILABILITY

The data that support the findings of this study are available from the corresponding author upon reasonable request.

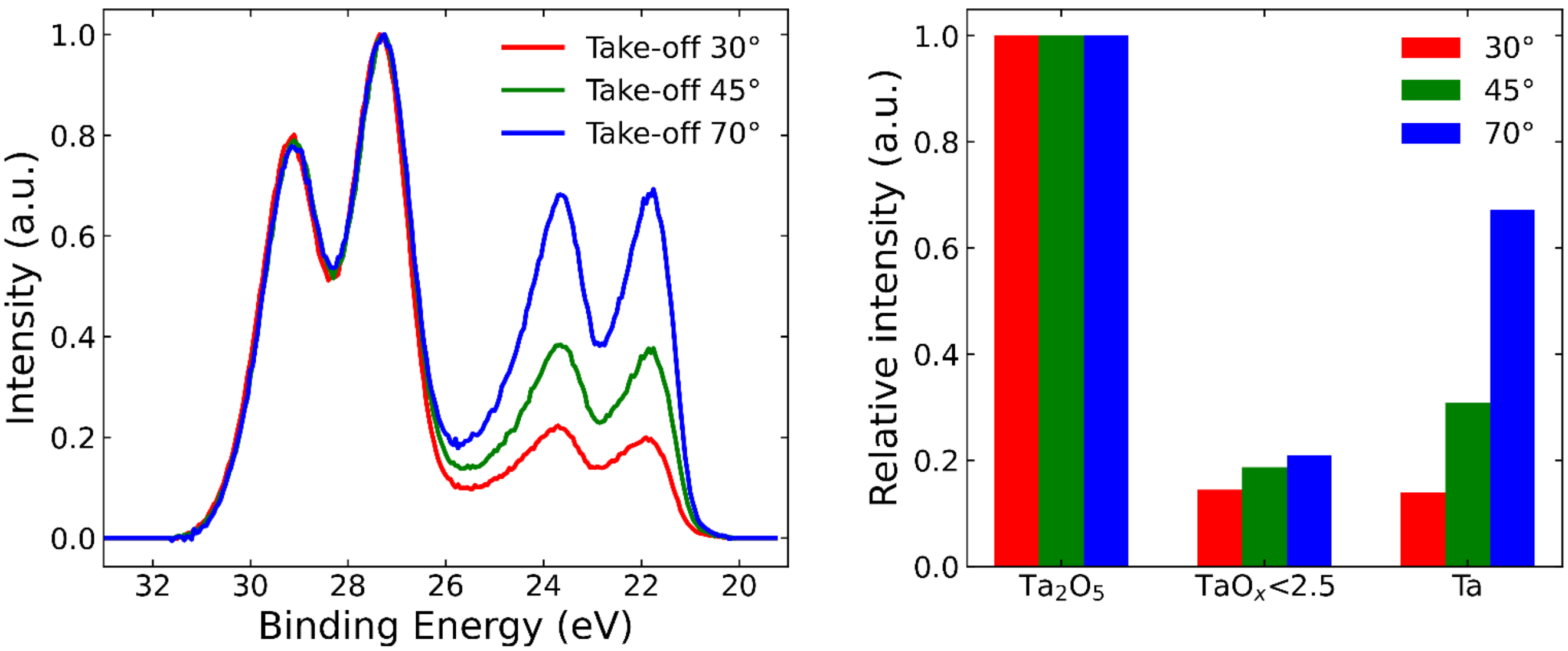

**FIG. 1.** Angle-resolved X-ray photoelectron spectroscopy (ARXPS) analysis of Ta oxidation states as a function of sampling depth. (a) Ta 4f core-level spectra acquired at take-off angles of 30°, 45°, and 70°, normalized to the $Ta_2O_5$ peak at each angle to show comparison of underlying oxidation states. (b) Relative peak intensities of $Ta^{5+}$ ($Ta_2O_5$), $TaO_x$<2.5, and $Ta^0$ extracted by CasaXPS curve fitting for each angle. The inset schematically illustrates the corresponding layered structure based on ARXPS analysis.

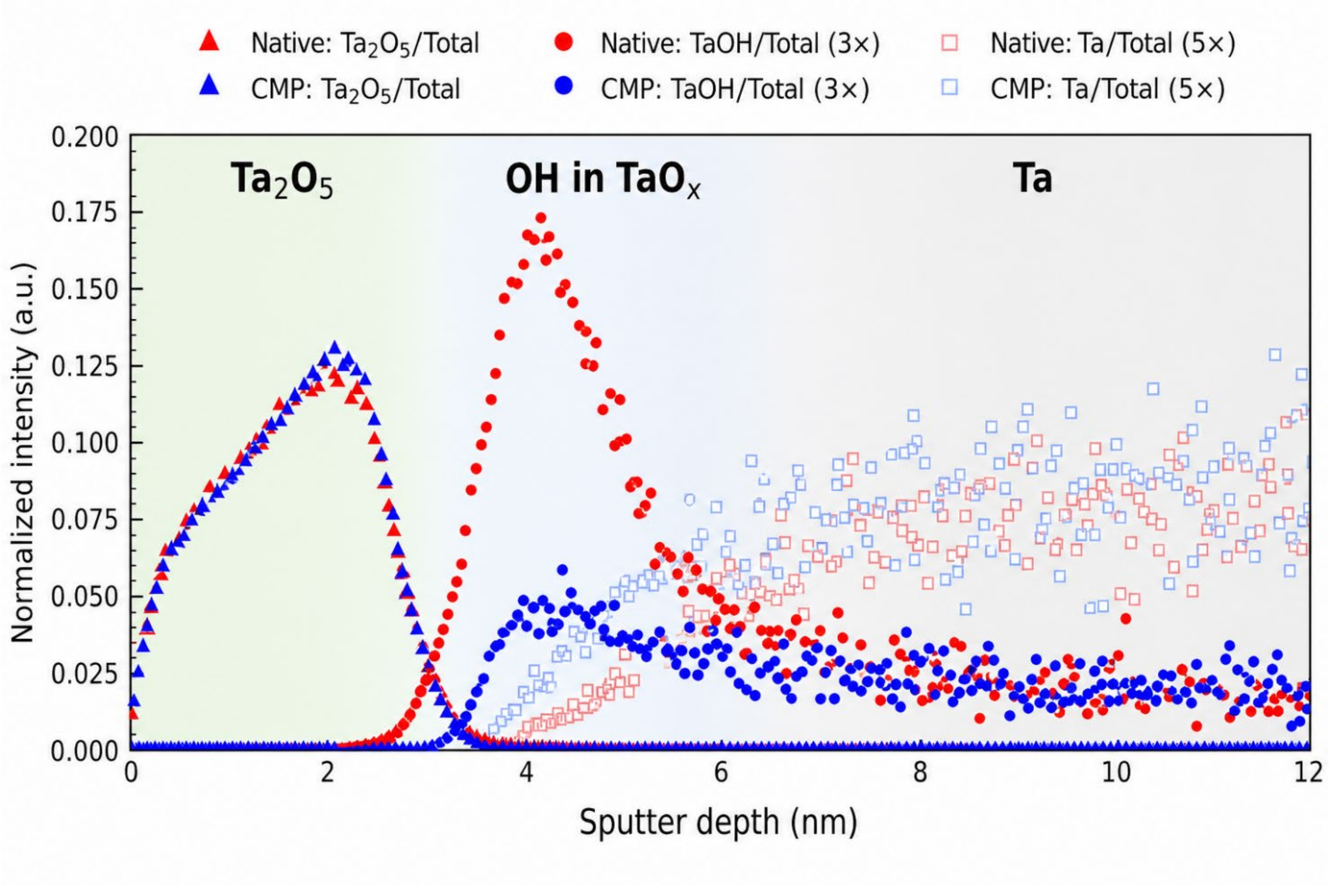

**FIG. 2.** Depth profiles of oxide species and hydroxyl content in native and CMP-formed tantalum oxides: SIMS data showing the normalized intensities of $Ta_2O_5$, TaOH, and Ta signals plotted as a function of sputter depth. The TaOH/Total signal reaches a maximum in an intermediate region beneath the $Ta_2O_5$ layer, this region is assigned to a $TaO_{x<2.5}$ suboxide based on the XPS analysis. For clarity, the TaOH/Total and Ta/Total signals are scaled by factors of 3× and 5×, respectively.

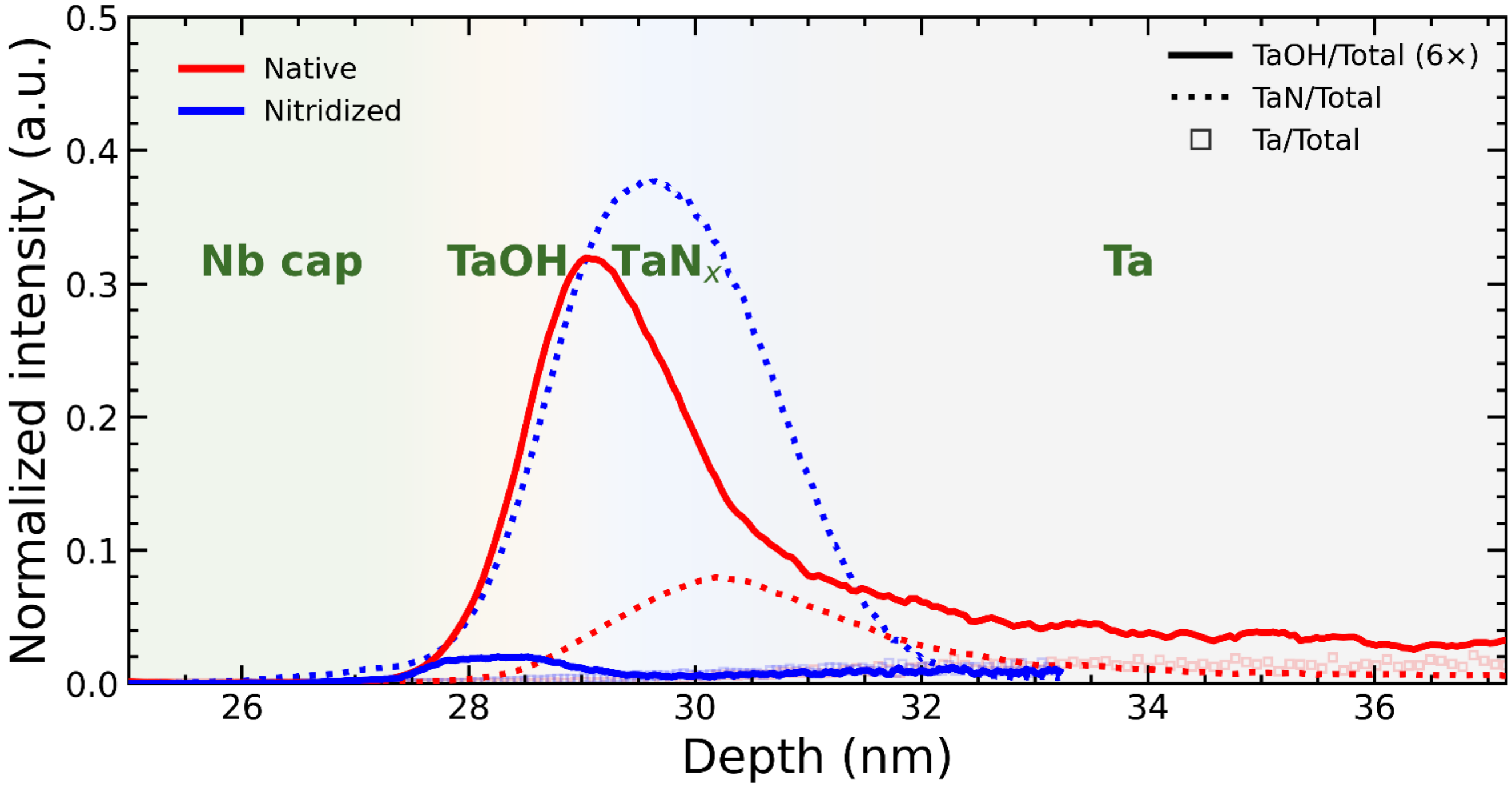

**FIG. 4.** SIMS depth profiles of hydroxyl content and nitridized Ta for nitridized samples capped after 3 weeks of ambient exposure and native oxide after 3 weeks of air exposure of Ta surface. Shown are normalized intensities (signal/Total) of $TaOH^-$, $TaN^-$, and Ta plotted versus sputter depth. The $TaOH^-$ signal reaches a maximum in the near-surface region above the $TaN_x$ layer. For clarity, the TaOH-/Total signal is scaled by 6×.

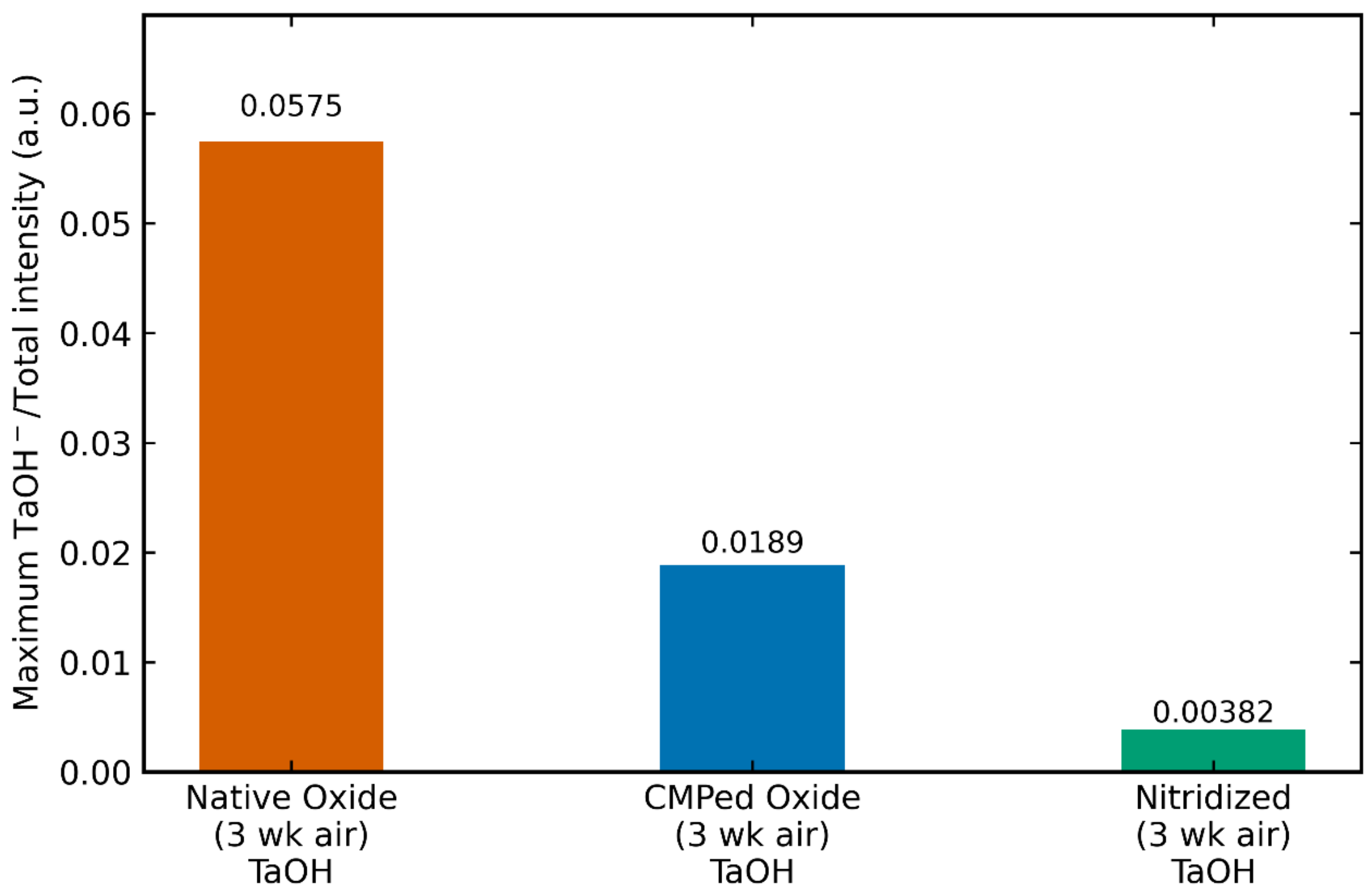

**FIG. 5.** Maximum $TaOH^-$/Total intensity for native oxide, CMP oxide, and nitridized/capped samples after 3 weeks of ambient exposure. Values were obtained from the maximum normalized SIMS $TaOH^-$ signal within the depth window corresponding to the hydroxyl-rich region in each case.